\documentclass[anonymous=false,sigconf,nonacm=true]{acmart} 

\AtBeginDocument{%
  \providecommand\BibTeX{{%
    \normalfont B\kern-0.5em{\scshape i\kern-0.25em b}\kern-0.8em\TeX}}}

\usepackage[utf8]{inputenc}
\usepackage{blindtext}
\usepackage{subfiles}   
\usepackage{graphicx}
\usepackage{array}

\title[Initiative and Materiality]{Initiative and Materiality: Exploring Mixed-Initiative Calculators\\ with the Tangible Human-A.I. Interaction Framework}

\author{Lunshi Zhou}
\affiliation{
\institution{Cornell University}
  \city{Ithaca, NY}
  \country{USA}}
  \email{lz489@cornell.edu}

\author{Alexandra Bremers}
\affiliation{%
  \institution{Cornell Tech}
  \city{New York, NY}
  \country{USA}}
\email{awb227@cornell.edu}

\author{Wendy Ju}
\affiliation{%
\orcid{0000-0002-3119-611X}
  \institution{Cornell Tech}
  \city{New York, NY}
  \country{USA}}
  \email{wendyju@cornell.edu}

\date{2024}


\begin{document}

\begin{abstract}
How can interactions with A.I. systems be designed? This paper explores the design space for A.I. interaction to develop tools for designers to think about tangible and physical A.I. interactions. Our proposed framework consists of two dimensions: initiative (human, mixed, or machine) and materiality (physical, combined, or digital form). A particularly interesting area of interactions we identify is the quadrant of \textit{physical, machine-initiated interactions.} With our framework, we examine calculator interactions and attempt to expand these to the tangible, mixed-initiative space. We illustrate each area in our proposed framework with one representative example of a calculator -- a common and well-known example of a computing device. We discuss existing examples of calculators and speculative future interactions with mixed-initiative and physical calculator systems. We reflect on the implications of our framework for the larger task of designing human-A.I. collaborative systems. Designers can also apply this framework as a guideline for analogous solutions to problems in the same domain.

\end{abstract}

\maketitle

\section{Introduction}
\begin{figure}
\makebox[0.5\textwidth]{\includegraphics[width=0.5\textwidth]{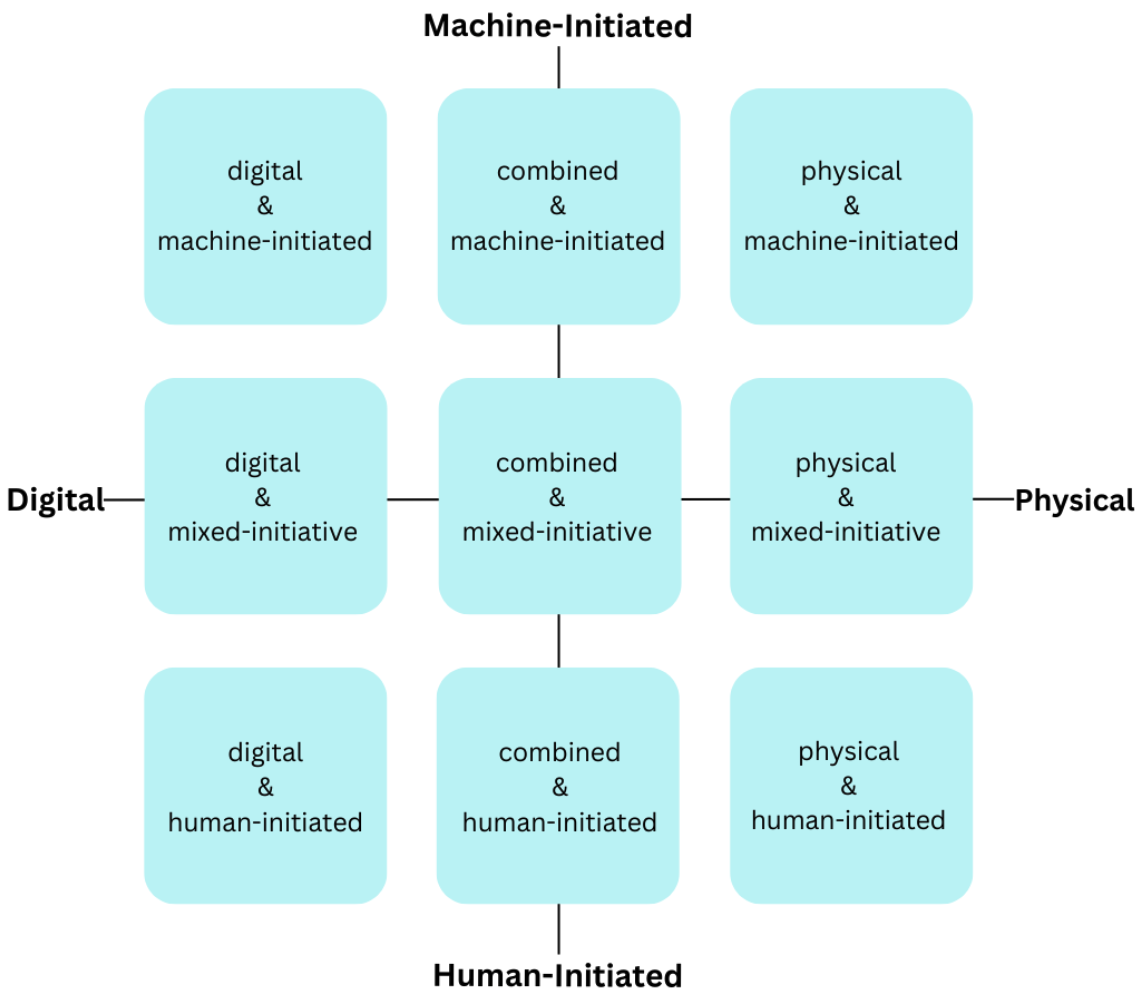}}
\caption{Schematic diagram of the Tangible Human-A.I. Interaction framework}
\label{fig_framework}
\end{figure}

Artificial Intelligence (A.I.) is a manufactured system that uses computations to perceive information and take actions to achieve goals \cite{AI_def2, AI_def1}. A.I. systems are usually good at specific tasks, yet they can struggle with complex environments -- designing smooth interactions with A.I. is still a challenge \cite{amershi2019guidelines,blackwell2015interacting}. 
A.I. applications can be either wholly digital or involve a physical interaction component, such as when computation is embodied in a robot. From years of study in tangible and embodied interaction, it is known that materiality influences interaction on many levels \cite{Physicality}. For instance, when robots are collaborative, influence, order, and space must be negotiated. Some recent works have investigated the potential for non-anthropomorphic robots to communicate initiative through the movement of their physical body \cite{ariccia2022make,grinberg2023implicit}.

In this late-breaking report, we approach how A.I. systems could be designed by developing a framework that can serve as a design tool (Figure \ref{fig_framework}). Our proposed framework spans the two dimensions of materiality and initiative. Initiative and materiality highlight this topic's relevance to human-robot interaction. We explore and illustrate the framework by reflecting on the design of a specific, extensively developed type of computational application, namely calculators. Calculators are the foundations or one of the most well-studied manifestations of computing systems, and they provide a concrete example to explore the more abstract design space of future A.I. systems. 
The examples of calculators mentioned in the framework are not intended to provide a systematic literature review but to illustrate the design space's utility. 
This paper contributes a framework to examine and design human-A.I. interactions that integrate initiative and materiality.

\section{Related Work}

\subsection{What are design spaces?}
The theoretical concept of a "design space" was first introduced to the realm of design research by Herbert Simon's Sciences of the Artificial, when he mapped the computer science notion of a problem or search space, for example, of all the moves that can be made in a chess game. This notion was adopted rather literally by researchers incorporating A.I. to perform design space exploration for computer-aided design in the 1990s \cite{esbensen1996design,esbensen1996design,moya1999evaluation}. In HCI and HRI, the notion of design space was interpreted more abstractly, for example, as the space of possibility and rationale embodied in early design prototypes that go beyond the mere specification or artifact or the realm of questions, options, and criteria being considered to approach a design problem \cite{maclean1991questions,bellotti1995multidisciplinary}. Nowadays, the conceptualization of design space is more metaphorical. Gaver, for example, describes "the notion of a 'design space'" as a metaphor for the designer's perception of possibility. He argues, "Design \textit{creates} the spaces in which it operates. They do not pre-exist their manifestation, whether as undiscovered design ideas or as the abstract and rationalized parameter spaces that some of those who follow Simon might suggest" \cite{gaver2011making}.

\subsection{Frameworks for the design of tangible interactions with A.I.}
Various relevant frameworks have been proposed to consider the design of interactions. For instance, \citet{ju2008design} presented the Implicit Interaction Framework which considers the dimensions of initiative and attentional demand; \citet{hornecker2006getting} presented a framework that integrates four different perspectives on designing tangible interactions, including a focus on social aspects of such interactions; \citet{mazalek2009framing} provides a meta-analysis of the frameworks that have been designed around the topic of tangible interactions and conclude that frameworks tend to focus on design and have less focus on knowledge abstraction or the technical development of interactive applications. Some prior work has looked explicitly into the promise of physicality for making A.I. more "graspable", using the form as a modality for explainability \cite{ghajargar2021explainable,ghajargar2022graspable}.

\subsection{A design space for A.I. interactions}
We believe there are numerous aspects of human-A.I. interaction where differences in the initiative, embodiment, and other interaction parameters create different genres of interaction, which generalize across applications. For example, in the space of human-autonomous vehicle interaction, the sensing and prediction capabilities of an "intelligent vehicle" could be used to act as a chauffeur for people, or it could be used to augment the driver's capability as a guardian, as illustrated in \cite{hodson2016your}. The Chauffeur/Guardian dichotomy can be useful for people working far outside vehicle interaction. To this end, our exploration of interactions with calculators is intended to be a foundation for potential directions of the development of human A.I. interactive systems and a way to make the abstract design space of A.I. systems more concrete.

\section{The Tangible Human-A.I. Interaction Framework}

\subsection{Framework dimensions} 
Figure \ref{fig_framework} shows two dimensions that we consider for human-A.I. interaction: 
\begin{itemize}
    
    \item \textit{Initiative (human-initiated/machine-initiated):} Does initiative lie with the human, the machine, or with both? If a human uses a typical calculator to get an answer to their mathematical question, the interaction is human-initiated. However, if we consider an intelligent program that can test users' progress in math, then the interaction is machine-initiated. Moreover, a third category exists of mixed-initiative interactions, where the users and machine collaborate \cite{mixed-initiative}. In our framework, we thus divide the vertical dimension into machine-initiated, mixed-initiative, and human-initiated interactions. 

    \item \textit{Materiality (physical/digital):} 
    Some calculators are totally programmed to run on computers, like WolframAlpha, which we would classify as digital. On the other hand, ancient calculators that do math computations purely on mechanical parts would be classified as physical. Here again, the dimension covers a spectrum between the two extremes. Many calculators have both mechanical parts and digital parts. 
    Therefore, we also divide the horizontal into three areas: from left to right, purely digital, combined digital and physical, and purely physical interactions.
   
\end{itemize}

\begin{figure}
\makebox[0.5\textwidth]{\includegraphics[width=0.5\textwidth]{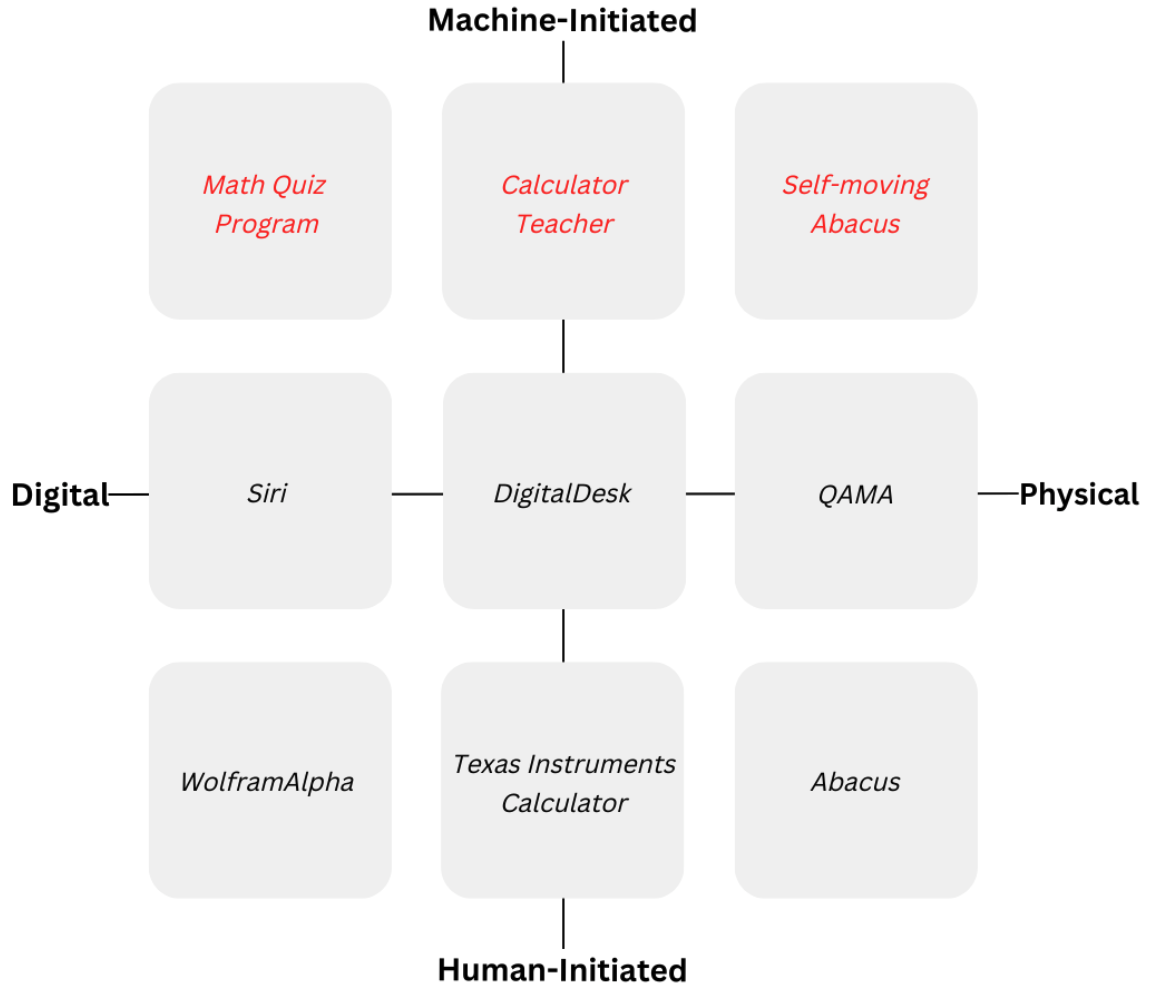}}
\caption{Schematic diagram of the Tangible Human-A.I. Interaction framework when applied to the case of calculators, with an example given in each grid space. Red fonts (top row) indicate three speculative examples.}
\label{fig_frameworkexample}
\end{figure}

\subsection{Example calculator interactions}

To better understand the range of the framework, we pick up existing examples and discuss people's interactions with these calculators. Figure \ref{fig_frameworkexample} shows examples we choose in each domain:

\begin{itemize}
    \item \textit{Abacus (physical \& human-initiated):}
    The abacus is an ancient calculator in human history that emerged before the adoption of the Hindu-Arabic numerical system \cite{Abacus}. Figure \ref{fig_Abacus} shows that it usually looks like a frame with beans and sticks connecting beams. This falls into the domain of physical, human-initiated interaction. People have developed many tricks for using the abacus as a supplementary tool to increase calculation speed. 
    \begin{figure}[t]
    \makebox[0.5\textwidth][c]{\includegraphics[width=0.25\textwidth]{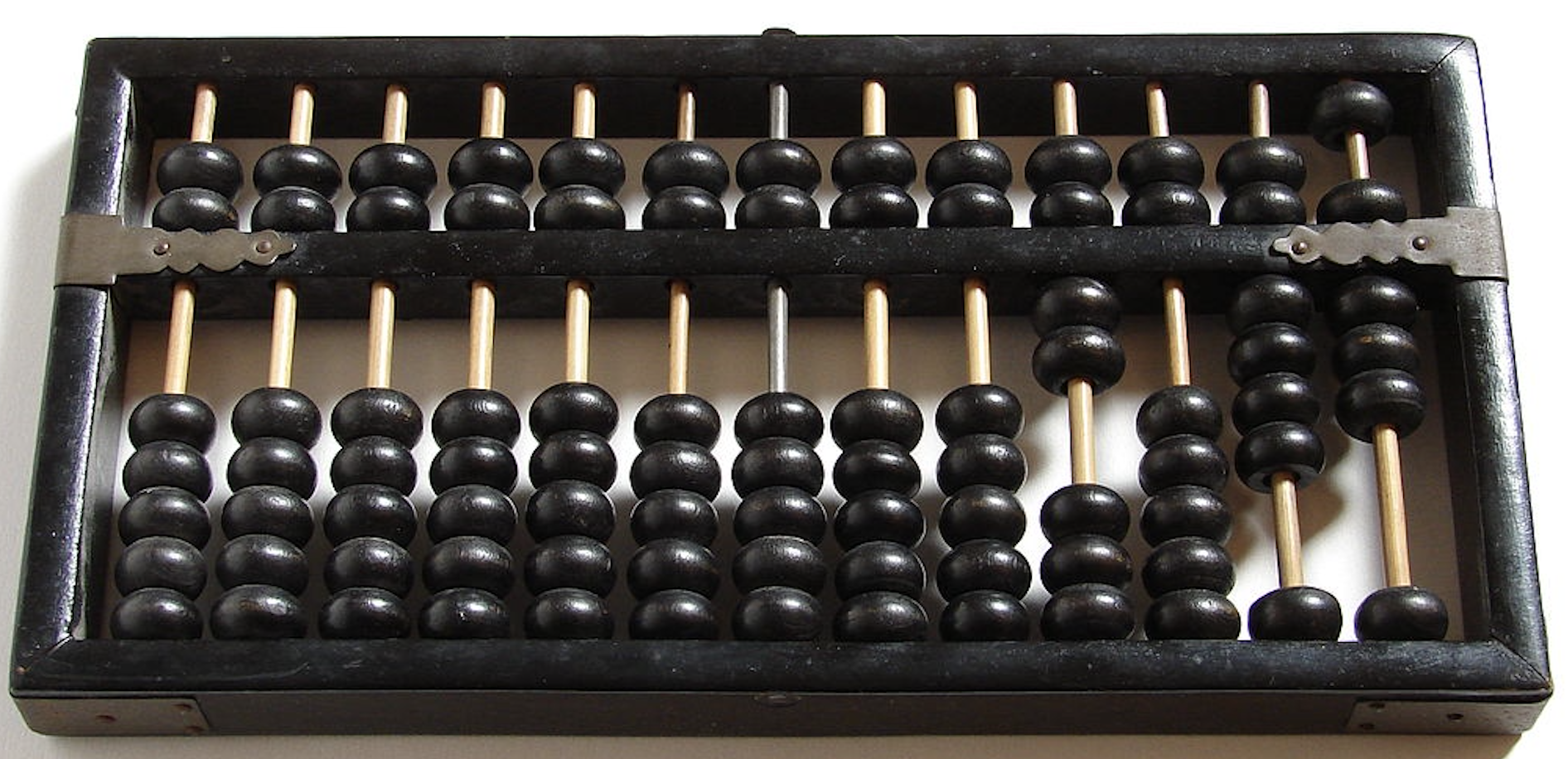}\includegraphics[width=0.12\textwidth]{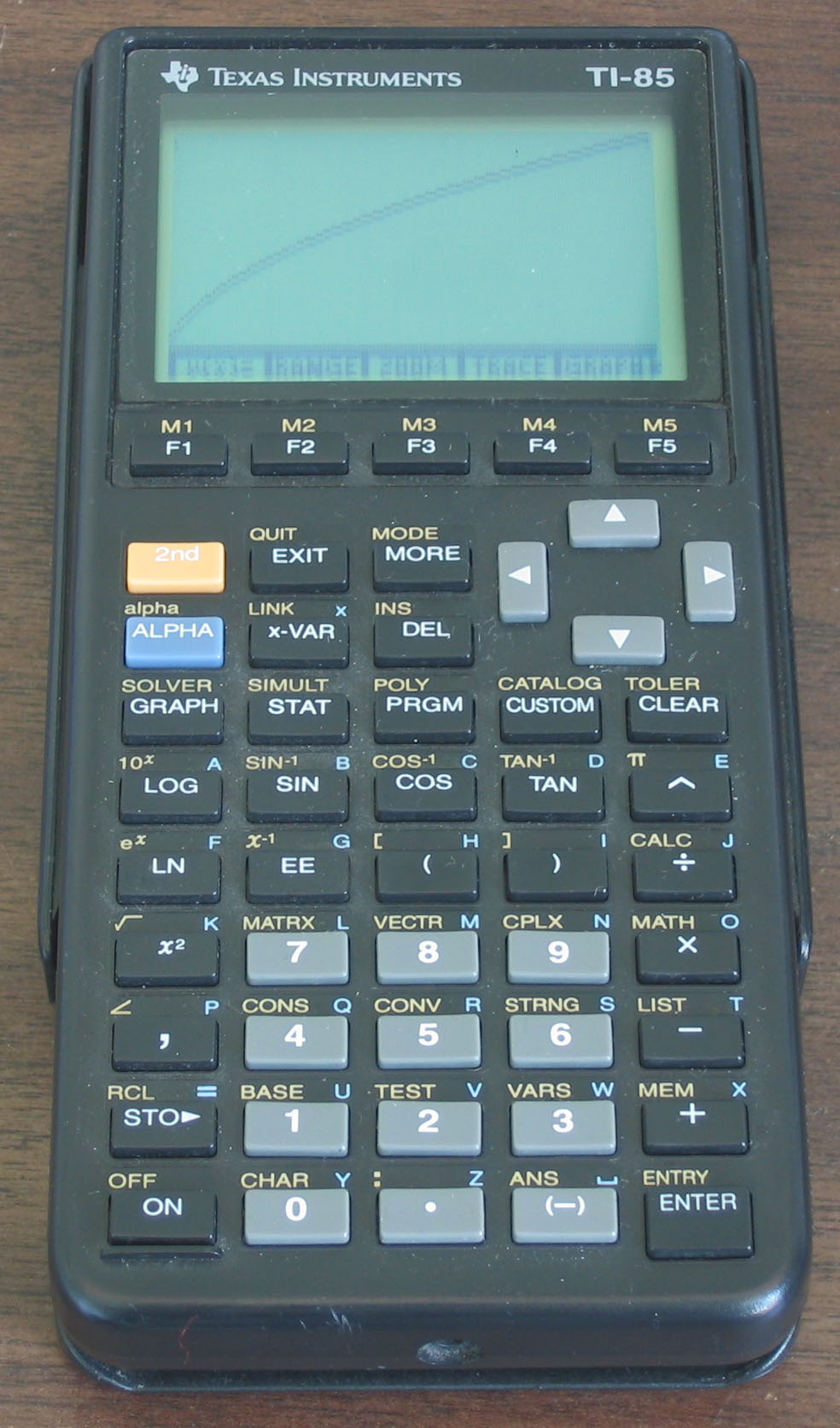}}
    \caption{Calculating using abacus\textsuperscript{1}, \textit{left}, Texas Instruments Calculator\textsuperscript{2}, \textit{right}}
    \small\textsuperscript{1,2} \textit{Image source: Wikimedia}
    \label{fig_Abacus}
    \end{figure}

    \item \textit{Texas Instruments standard calculator (combined materiality \& human-initiated):}

    Texas Instruments calculators are representative of classic mechanical calculators with physical keys to input math symbols and a screen to demonstrate results (see Figure \ref{fig_Abacus}). Hence, its materiality has a combination of physical and digital parts. The calculator heavily depends on the user's input and does not take initiative independently. People mainly use such a category of calculators as an essential tool to support complex calculations, like multi-digit multiplications, that are time-consuming or require great precision. It appears in a wide range of engineering scenarios to provide help with calculations.

    \item \textit{WolframAlpha web calculator (digital \& human-initiated)}:

    WolframAlpha is a web-based calculating engine that provides solutions to a wide range of Math topics (see Figure \ref{fig_WolframAlpha})\cite{WolframAlpha}. Users mainly use it through web browsers. The users can input math symbols through the mouse by clicking the desired functionalities listed on the WolframAlpha page. It can demonstrate answers and steps to reach answers on the computer. Therefore, such a calculator is categorized as digital and mixed-initiative. People mainly use WolframAlpha as a tool to do complicated calculations beyond the scope of mechanical calculators and default calculators in computers.

    \begin{figure}
    \makebox[0.5\textwidth]{\includegraphics[width=0.45\textwidth]{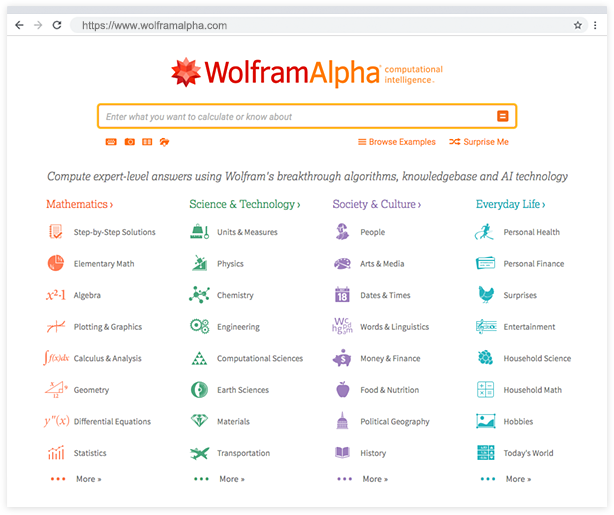}}
    \caption{WolframAlpha Main Page\textsuperscript{3}}
        \small\textsuperscript{3}\textit{Image source: www.wolframalpha.com}
    \label{fig_WolframAlpha}
    \end{figure}

    \begin{figure}
    \makebox[0.5\textwidth]{\includegraphics[width=0.45\textwidth]{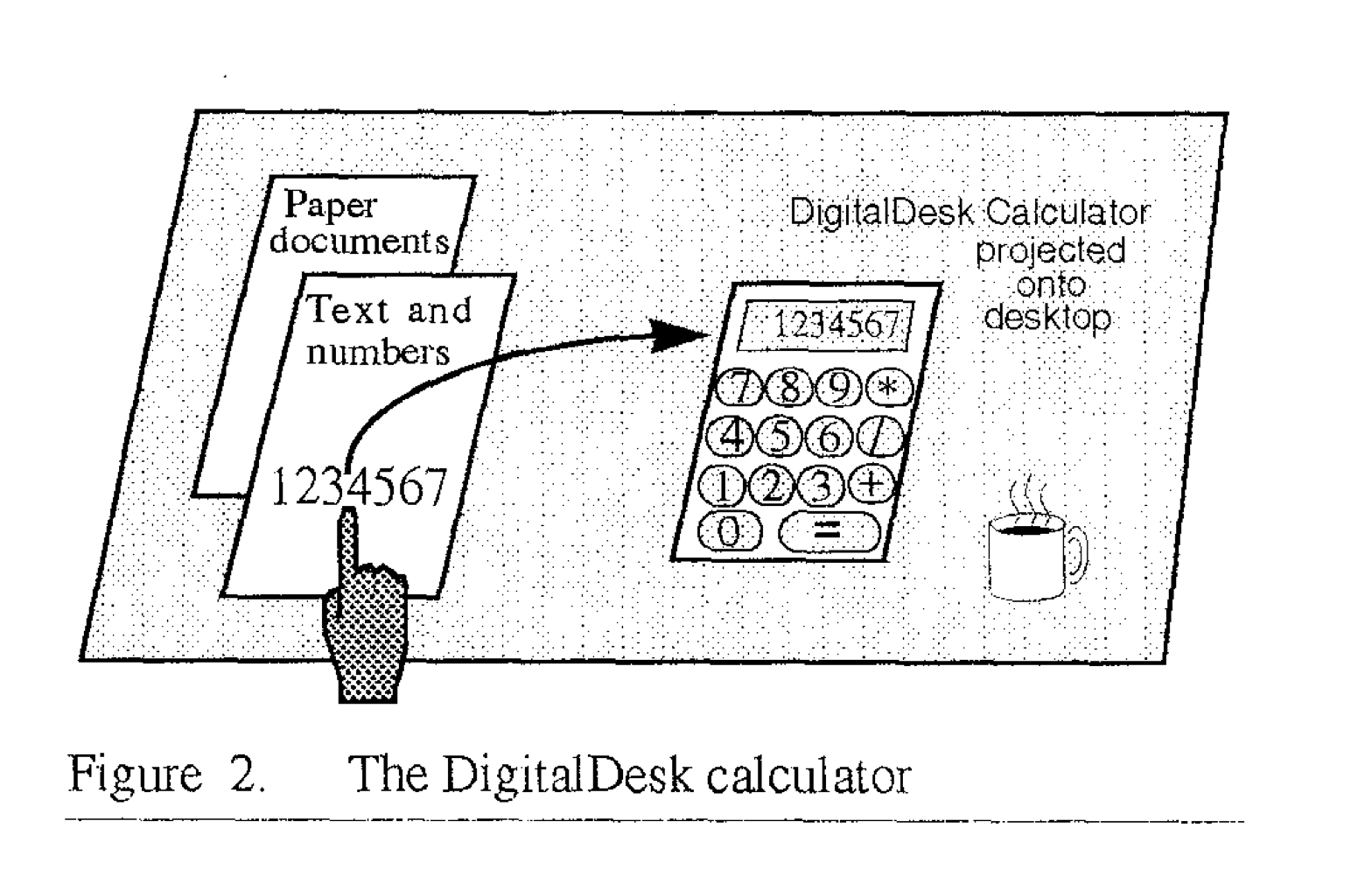}}
    \caption{Digital Desk by Pierre Wellner\textsuperscript{4}\cite{DigitalDesk}}
        \small\textsuperscript{4}\textit{Image reproduction permission granted by Wellner}
    \label{fig_DigitalDeskCalculator}
    \end{figure}

    \item \textit{Siri (digital \& mixed-initiative):} Siri is a virtual assistant developed by Apple that can perform calculations based on spoken inputs, with the possibility to ask clarifying follow-up questions \cite{Siri}. For example, a user can ask, "What is 1 plus 1?" and Siri will answer, "The answer is 2". Siri can also perform complicated calculations by using internet resources. People can use Siri when they have a hard time using their hands to access other calculators, such as when a mechanic is busy with their hands but needs to do quick math for a price quote. Since the user needs to give verbal instructions for the calculations and Siri can perform them independently, we classify it as an example of a mixed-initiative interaction. As Siri is purely digital, as software is installed in Apple products, we say it is a digital and mixed-initiative calculator. 
    
    \item \textit{Pierre Wellner's digital desk (combined materiality \& mixed-initiative):}
    Pierre Wellner developed a prototype of a DigitalDesk that allows users to interact with paper and electronic objects by touching them with a bare finger on a physical desk equipped with a computer-controlled camera and projector above it (see Figure \ref{fig_DigitalDeskCalculator}).
    \cite{DigitalDesk}. Since this prototype allows users to interact directly with physical and digital materials, we classify it as a combined format. And because of its ability to recognize physical manipulations and produce corresponding digital changes under bare-finger manipulations, we classify it as a mixed initiative.

    \item \textit{QAMA (physical \& mixed-initiative):}
  
    QAMA is a physical calculator that shows the answer only if a user provides a precise enough estimate \cite{QAMA}. Its target is students who are studying math concepts, aiming to help students better understand math concepts through mental estimates. QAMA uses algorithms to enable the calculator to decide which estimations are precise enough and hence, evaluate whether the student masters the concept. Based on this, QAMA is a mixed-initiative calculator with a physical body since it still requires users to input equations actively. 

\end{itemize}

\section{Discussion}
\subsection{Evaluating the framework}

Frameworks are like models--as the saying goes, all models are wrong, and some models are useful. The evaluation of our framework, then, should be based on its utility. The exploration of calculators in the framework serves as an example of the application of other forms of computing systems. Since the dimensions considered in this framework are pertinent to all forms of A.I., we believe the framework could inspire designers to find analogous solutions across different A.I. systems. For example, there may be some common ground in interactions between an A.I. program that teaches people to drive in virtual reality and an A.I. app that teaches students math. There can be shared features across such A.I. systems, and designers can look for inspirations or solutions across different systems in the same framework domain. In the next stage of this project, we are looking to refine and evaluate our framework by including interview data from speaking to users and practitioners in the space of A.I. system design, using our framework and certain specific instances of its application as design probes.

\subsection{Future Work: Trying out three modes of mixed-initiative and machine-initiated calculators with Wizard-of-Oz}
Through our framework development and exploration of the design space of calculator interactions, we included six examples of interactions that are already fully developed products or prototypes. The domains of digital, physical, and human-initiated interactions seem to offer plenty of examples of interactions that have been investigated and fine-tuned for decades. 

Importantly, we identified key areas of the framework that are largely gaps in human-A.I. interactions, namely machine-initiated and mixed-initiative interactions, particularly when physical materiality is involved. Since there are no existing examples yet in some spaces, we plan to explore these areas through a case study in which we will design mixed-initiative interactions with a calculator. We will consider three designs that cover the three types of materiality in our framework:

\begin{enumerate}
\item \textit{Digital:} Such a calculator can have the form of an app installed on computers or smartphones. It can have an A.I.-controlled program that reminds students to do math based on their schedules and sets the difficulty of the questions based on previous answers of students.
\item \textit{Combined materiality:} This kind of calculator can have a combined format of physical and digital content. We propose a physical calculator equipped with screens and cameras, which enables it to read students' facial expressions to determine whether they are confused about questions it asks students to compute. If so, it will demonstrate detailed calculation steps and guide students in following them. If not, it will simply show answers.
\item \textit{Physical: }This type particularly emphasizes the physical component. Based on this principle, we propose a self-moving abacus with a screen to demonstrate some calculation techniques and can self-move beans to simulate student situations. 
\end{enumerate}

We will evaluate the three calculator interaction concepts through a user study with 18 participants, where users are video-recorded. At the same time, they perform calculation tasks with the calculators in counter-balanced order. The participants consist of designers as well as users of computer systems. After the task, the users will be asked to complete quantitative questionnaires to rank the various calculator designs. They will participate in a semi-structured qualitative interview where they are asked about their experience and how it makes them think about using or designing for A.I. systems. For setting the interaction, we plan to rely on interaction prototyping and Wizard-of-Oz systems, such as described in \cite{dahlback1993wizard,hoffman2014designing,riek2012wizard}.

\section{Conclusion}
In this late-breaking report, we explore human-A.I. interaction through a proposed framework of two dimensions: initiative and materiality. These are pertinent elements across the design of different A.I. systems. This paper further provides illustrative examples of calculators for six areas that have already been well-developed with existing calculators and identifies areas of the framework with no existing calculator examples, notably machine-initiated or mixed-initiative interactions. Future work will investigate the application of the framework to concrete systems tested with users and practitioners to understand the value such a framework could offer designers. With our illustrated framework, we hope designers can look for analogous solutions or gain inspiration across different A.I. systems in the same domain. 

\section{Acknowledgements}
We would like to thank members of the Cornell Tech research community for their suggestions and ideas regarding this project.

\bibliographystyle{ACM-Reference-Format}
\bibliography{bibliography,designspace,newrefs}

\end{document}